# Regeneratives Vibrations Analyzes On The Part And Chip

CONSTANTIN ISPAS[1], CLAUDIU BISU[1], ALAIN GERARD[2], DORU BARDAC[3]
[1] Machines and Production Systems Departament
"POLITEHNICA" University of Bucharest
Splaiul Independenţei no. 313 Street,
ROMANIA
[2] Mechanic and Physic Laboratory
University Bordeaux 1
351 Cours de la Liberation, Tlaence 33405 Cedex
FRANCE
[3] Manufacturing Department
"POLITEHNICA" University of Bucharest
Splaiul Independenţei no. 313 Street,
ROMANIA
ispas1002000@yahoo.fr, cfbisu@gmail.com, alain.gerard@u-bordeaux.fr, doru@bardac.net

*Abstract:* - The actions of cut applied to the elastic system generate relative tool/part displacements, which induce a temperature increasing in the components of the machine tools, its environment, the system-tools-part and generate vibrations. The experimental procedures used, at the dynamic level, made it possible to determine the elements necessary to analyze the influence of the tool geometry, its displacement and evolution of the contacts tool/part and tool/chip on surface carried out.

## 1 Introduction

The dynamic phenomena of the machine tools come from the interaction of the elastic system machine-process of cut. This interaction is thus the generating source of the dynamic aspects classically met in the machine tools. The cutting actions applied to the elastic system cause relative displacements tool/part, which generate variations. Those modify the chip section, the contact pressure, the speed of relative movement etc. Consequently, the cutting process instability can cause the dynamic system instability of the machine tool vibrations appear. They have reflected undesirable on the machined surfaces quality and the tool wear. They can generate problems of maintenance even ruptures of machine tools elements. Thus, it is necessary to develop models making it possible to study the vibratory phenomena met during machining and to envisage the stable conditions of cut.

The effect has been established important ace one off the most sources off sustained relative vibrations between the tool and the workpiece and has been extensively studied in the literature [1], [2], [4].

Considerable efforts are developed to model the cut correctly. But at the present time the solutions suggested are still far from providing sufficiently relevant and general models to translate in an acceptable way the unit of the experimental results available. Taking into consideration cutting condition three-dimensional and nonstationary, in particular in the case of the vibratory modes, [6], [7], the models to be implemented becomes.

## 2 Experimental Device

The dynamic cut tests are carried out on a conventional lathe (fig.1). The behavior is identified with a three-dimensional accelerometer fixed on the tool and two unidimensional accelerometers positioned on the lathe, side stitches, to identify the influence of the pin on the process. The efforts and the moments, expressed at the tool tip, are measured using a six components dynamometer. The evolution the instantaneous speed of the part is given by a rotary coder. The three-dimensional dynamic character is highlighted by seeking the various existing correlations or the various evolutions of the parameters, which make it possible to characterize the process.

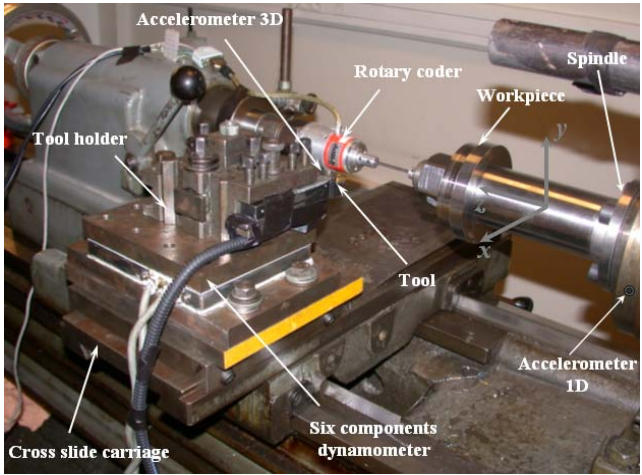

Fig.1. Experimental device

This analysis provides the necessary information to the design and the development of model of the three-dimensional dynamic cut. Thus, the experimental study of the phenomena met is essential to obtain a model reflecting the processes reality present at the cut time.

The vibrations measurement during the machining led to the results presented to the. The cut vibrations frequencies are around 190 Hz for the three axes with maximal amplitude on the $y$ axis (cut axis), fig 2 and fig.3. We consider during the analysis that the recorded frequencies correspond well to the regenerative vibrations related to the cut phenomenon. The frequency peack around 200 Hz is found for the other tests according to the feed rate. The regenerative vibrations have an influence on the surface quality of the parts. Analysis FFT of the rugosimetric data shows a frequency peak located around 190 Hz, which is coherent with the preceding data.

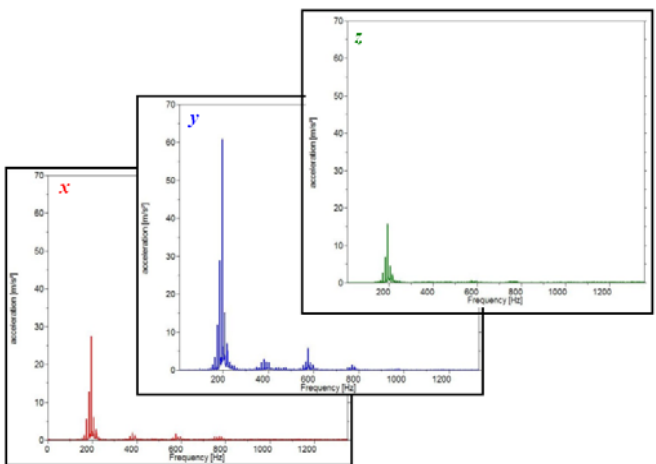

Fig. 2 The FFT acceleration signal in the three direction cutting

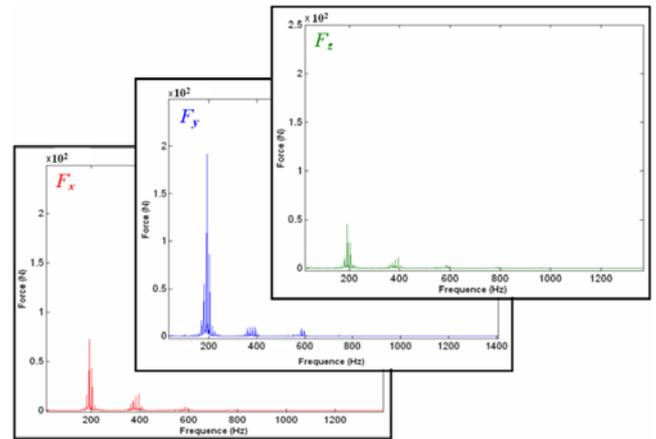

Fig. 3 The FFT forces signal in the three direction cutting

## 3 Part And Chip Geometry

The regenerative vibrations have an influence on the surface quality of the parts (fig.4). Analysis FFT of the rugosimetric data shows a frequency peak located around 200 Hz, which is coherent with the preceding data. on the chips and made it possible to determine the variations thickness and width of those.

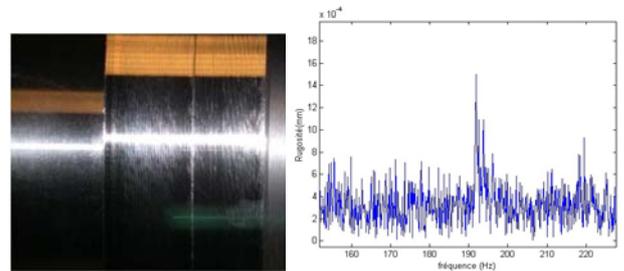

Fig.4. FFT of the roughness profile of the machined part.

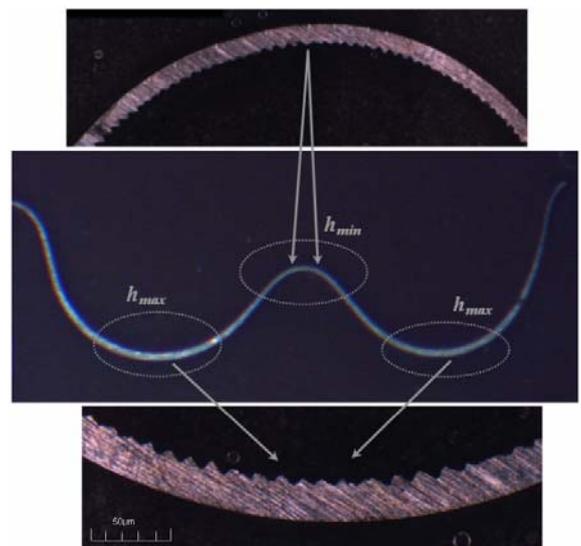

Fig.5. Variation chip thickness.

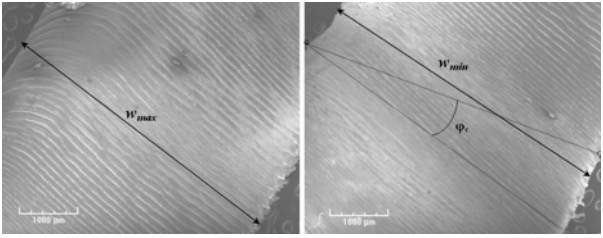
Fig.6. Variation of the chip width.

An example is presented (fig.5) for a sample of chip during a test with an advance of 0.05 mm/tr where $h$ =0.23mm and $h$= 0.12mm. The measure of chip length corresponding to an undulation enables us to find the frequencies of self-sustained vibrations starting from the cutting speed. To determine the overall chip length, it is necessary to measure the wavelength and to take account of the chip rate at the cut time.

$$f_{cop} = \frac{V}{l_o \cdot \xi_c}, \quad (1)$$

with $f_{cop}$ the chip frequency, $V$ the chip speed, $l_o$ the undulation length of chip and $\xi_c$ the work hardening coefficient for the chip. The coefficient of work hardening is given using the it is then validated by the literature.

$$\xi_c = \frac{\cos(\phi-\gamma)}{\sin(\phi)}, \quad (2)$$

In our case, 11 mm length of undulation is measured on the chip, with a coefficient of work hardening $(\xi_c)$ from 1.8 and one cutting speed $(V)$ of 238 m/min. We obtain then, a frequency of 206 Hz, very near to thee frequencies of displacements or forces raised at the time of the cut. The chip width is then measured with the same techniques. Important variations, about 0,5mm, are observed. The maximum width $(w_{max})$ is of 5.4 mm and the minimal width $(w_{min})$ 4.9 mm, (fig.6). The measurement of the slopes angle (fig.7) of the chip width between each undulation, close to 29°, is equal to the dephasing measured on the displacements signals (fig.7, fig.8 and fig.9 ).

## 4 Regenerative vibrations, Experimental validation

According to this study, the zone of self-excited vibrations is considered around 190 Hz. The analysis carried out to the measures of displacements and the forces makes it possible to evaluate existing dephasing between forces and displacements (Table 1). This dephasing explains the delay of the force compared to displacement. The appearance of regenerative vibrations can be also explained by the delay forces/displacement, which increases the energy level in the system. The existence of this delay is explained by the inertia of the machining system and more particularly by the inertia of the cut process [1]. Dephasing forces/displacement remains constant according to the advance. During this process, work remains positive and the resulting energy stored by the system allows the vibrations maintenance [3]. The experimental procedures installation, at the dynamic level, made it possible to determine the elements necessary to a rigorous influence analysis of the tool geometry, its displacement and contacts evolution tool/part and tool/chip on surface carried out.

*Table 1*
Dephasing values forces/displacement.

| $\varphi_{fu\_x}$ | $\varphi_{fu\_y}$ | $\varphi_{fu\_z}$ |
|---|---|---|
| 13° | 23° | 75° |

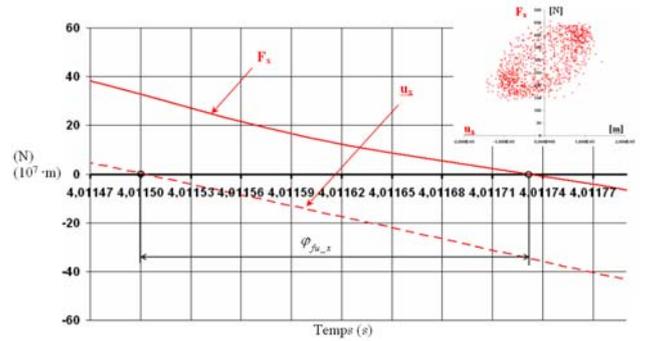
Fig.7. Dephasing forces/displacements according to $x$

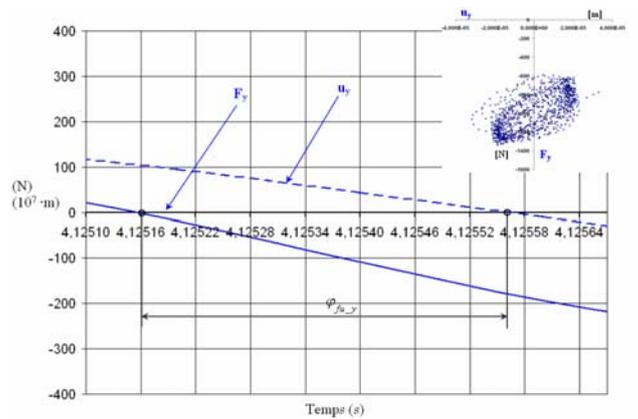
Fig.8. Dephasing forces/displacements according to $y$

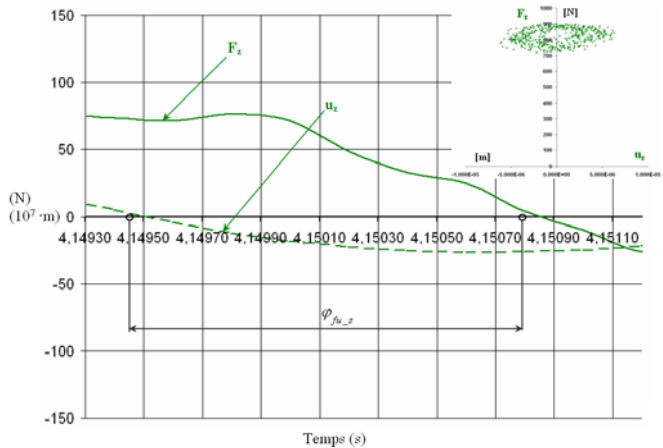
Fig.9. Dephasing forces/displacements according to *z*

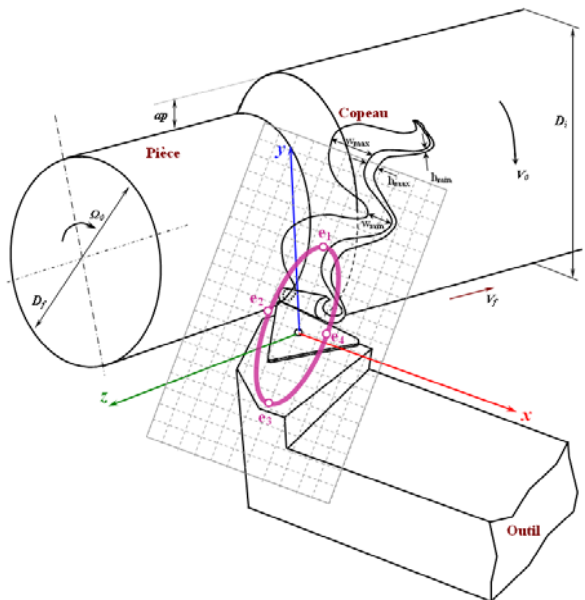
Fig.10. Elliptic trajectory of the movement the tool/part

## 5 Conclusion

The appearance of the self-sustained vibrations thus remains expliquable by dephasing between the forces and displacements [3]. When the tool moves on the part $e_1,e_2,e_3$ ellipse, the cutting force carries out a positive work because its direction coincides with the direction of cut. On the part $e_3,e_4,e_1$, the work produced by the cutting force is negative. By comparing the two parts of the ellipse, we notice that the force on the course $e_1,e_2,e_3$ is larger than on the course $e_3,e_4,e_1$ because the cut depth is larger (fig.10).

In addition, the coupling highlighted between the elastic characteristics of the machining system and the vibrations generated by the cut made it possible to that the self-sustained vibrations appearance is strongly influenced by the stiffnesses of the system, their report/ratio and their direction. We also established a correlation between the direction of the elastic structure vibratory movement of the machine tool and the variations of chip section. The self-sustained correlations part-chips/vibrations are valid. These first results make it possible to consider a more complete study by exploiting the concept of total mechanical actions. Indeed, thanks to the six components dynamometer, the existence of moments, not present here, is identified at the tool tip point. This side is not evaluated by the traditional measuring equipment. In the long term, the originality of work partly presented here relates to the analysis of the torque of complete actions applied to the tool tip point, with an aim of making evolve/move a model semi analytical 3D of the cut.